\documentstyle[11pt,psfig,vsopsymp]{article}
\begin{document}


\markboth{\hfill Garrett et al.}{Deep VLBI Imaging \hfill}


\centerline{\LARGE\bf 
   A New Strategy for the Routine Detection } 
\bigskip

\centerline{\LARGE\bf \& Imaging of Faint Radio Sources
  with VLBI}
\bigskip

\centerline{\sc 
M.A.~Garrett
}\medskip

\centerline{\it
JIVE, Postbus 2, 7990 AA Dwingeloo, The Netherlands.  
}

\begin{abstract} \noindent 
  
  In this paper I outline a new strategy for the routine detection and
  imaging of faint (sub-mJy and microJy) radio sources with VLBI and
  SVLBI.  The strategy relies on a combination of in-beam
  phase-referencing, wide-field VLBI imaging and simultaneous
  correlation of multiple field centres.  A combination of these
  techniques, together with the steeply rising radio source counts
  observed at cm wavelengths, permit routine high resolution
  observations of radio sources previously considered too faint for
  conventional VLBI.

\end{abstract}


\sources{Sub-mJy sources}

\section{Introduction}

VLBI is sensitivity limited. Most sources that can be robustly detected
by conventional self-calibration techniques have peak fluxes in excess
of 10~mJy.  The success of phase-referencing techniques, as applied to
mJy and a few sub-mJy radio sources, are often limited (particularly in
terms of image fidelity) to the brighter sources for which subsequent
self-calibration (over much longer solution intervals) is then
possible.  So far, few attempts have been made to detect sub-mJy
sources, despite the fact that with a coherent integration time of 24
hours, global VLBI arrays can routinely produce images with $1\sigma$
rms noise levels better than $30\mu$Jy/beam.

Nevertheless, the focus of VLBI over the last 3 decades (and in
particular Space VLBI -- SVLBI) has been directed towards the study of
the brightest and most compact radio sources in the sky. At these flux
levels ($> 10$mJy), the radio sky is virtually empty, with most radio
sources associated with relatively distant AGN. As a result the overlap
with other wave-bands is sometimes limited. In this paper, I suggest a
new strategy for the {\it routine\/} detection and imaging of faint
sub-mJy and $\mu$Jy radio sources. The strategy relies on a combination
of in-beam phase-referencing (with obvious advantages for SVLBI but
also VLBI generally - see Fomalont et al.  1999), wide-field VLBI
imaging (see Garrett et al.  1999) and simultaneous correlation of
multiple field centres. These techniques, together
with the steeply rising radio source counts at $\lambda$cm wavelengths,
should permit high resolution, VLBI investigations of the faint sub-mJy
and microJy source populations to begin.

\section{Towards Routine Imaging of Faint Radio Sources} 

It is a well known and auspicious fact that the radio source counts
increase steeply as one goes to fainter flux levels. At $\lambda$18cm
the source counts derived from WSRT observations of the Hubble Deep
Field, HDF, (Garrett et al. 2000) imply source counts of up to $\sim
40S^{-1}_{\mu Jy}$ per square arcmin. Thus within the central regions
of the primary beam of a typical 25-m VLBI antenna, one can expect to
find over $\sim 100$ potential target sources with $S > 120 \mu$Jy (the
$3\sigma$ noise level routinely achieved in ground based VLBI images).
If we extrapolate the preliminary results of Garrington, Garrett and
Polatidis (1999), we can deduce that for every continuum VLBI
observation conducted today, there are perhaps a dozen or so faint
radio sources in the beam that might be compact enough to be detected
and imaged, {\it in addition} to the brighter target source!  This
suggests a new strategy for the routine imaging of a large number of
faint radio sources:

\noindent
(i) {\it Reverse} the traditional approach of selecting the target
before the calibrator (see also Garrington, Garrett \& Polatidis 1999).
The field chosen should satisfy the following criteria: (a) it should
be an area for which high quality optical/IR, and deep, sub-arcsec
resolution radio data are available (several such fields are expected
to become available over the next year) and (b) the same field should also
contain a reasonably bright radio source (but not too bright)
that can act as a in-beam (secondary) phase-calibrator.

\noindent  
(ii) Split the antenna primary beam into manageable
$4^{\prime}\times4^{\prime}$ patches (the size of these patches is
currently determined by the limiting integration time and frequency
resolution provided by current generation correlators, not to mention
throughput, offline storage sizes and processing speed!). Each patch
can be generated via simultaneous multi-field centre processing
(currently being developed for the JIVE correlator, Pogrebenko 2000) or
standard multiple-pass correlation, and can share the phase corrections
provided by the in-beam phase-calibrator located close to the centre of
the beam.

\noindent  
(iii) Divide each phase calibrated (but unaveraged) data patch into many 
small sub-fields of a few arcseconds across (small enough to employ
2-D FFTs and not so large that the image size becomes unmanageable - at
least in terms of casual inspection by eye). FTT the data and produce a
dirty map of the sub-fields of interest. 

\begin{center}
\begin{figure}
  \vspace{70mm} \includegraphics{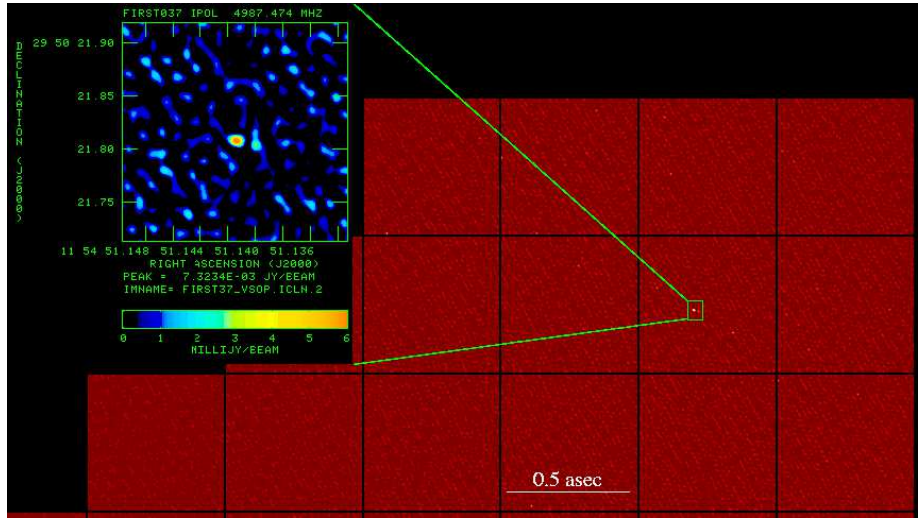}

\caption{Wide-field VLBI dirty image of a region of sky located 1 arcminute 
  from the phase-centre and target source. The detection of a
  previously known $\sim 10$ mJy VLA FIRST source in one of the sub-fields 
  is produced from only 10 min of (unaveraged) global 6cm VLBI data.}

\label{fig1}
\end{figure}
\end{center}

\vspace{-1cm} 
Some simple ``proof-of-concept'' tests have been conducted with a total
of 10 minutes of Global VLBI $\lambda$6cm data (taken from the
$\lambda$~6cm Global VLBI Faint Source Survey of Garrington, Garrett \&
Polatidis 1999).  Fig.~\ref{fig1} shows the clear VLBI detection of a
known VLA FIRST source in a sub-field that is part of a patch of the
primary beam that is located $\sim 1$ arcminute from the phase-centre
and target source. Details of the processing requirements for this (and
longer runs) is beyond the scope of this paper but they are not
unreasonable. A more important limitation is the minimum integration
time provided by today's working correlators (these are currently
inadequate to cope with SVLBI at Perigee - using this particular strategy
- but improvements can be expected over the next few years).

\section{The Structure of Faint Radio Sources \& SVLBI-2}

Exceptionally deep radio observations of the HDF (Richards et al. 1999,
Muxlow et al. 1999, Garrett et al. 2000) show that the bulk of the
sub-mJy and $\mu$Jy source population have steep radio spectra and are
for the most part identified with distant disk or irregular,
interacting galaxies (often with ISO detections). This argues strongly
that these faint sources are associated with very luminous Starburst
galaxies.  Nevertheless, a significant fraction (perhaps as much as
30\%) are probably faint AGN, especially the brighter sub-mJy sources.
Using the techniques described here, the brighter AGN could be
reasonable targets for the next generation of SVLBI missions now
planned. Indeed, SVLBI observations are probably crucial: from simple
SSA theory faint sources are also expected to be small. In addition,
emission from both compact AGN and larger-scale star-forming regions
(principaly young SNRs, relic SNR emission and ultra-compact HII
regions) might not be uncommon in the same system. Even for relatively
distant ($ \leq 350$~Mpc) but ultra-luminous star-forming disk
galaxies, hypernovae (such as those SNR in Arp 220 and 41.95+575 in
M82) might be detected, and more importantly {\it resolved} by SVLBI-2
missions. The prospects of detecting these faint, steep sepctrum radio
sources with next generation SVLBI missions depends crucially on the
availability of L or S-band receivers. The contribution future SVLBI-2
missions could make to unravelling the nature, structure and
composition of the faint radio source population cannot, and should
not, be underestimated.

\end{document}